\documentclass[9pt,conference]{IEEEtran}
\usepackage{amsmath,graphicx,url,times,booktabs, tabularx}
\usepackage{makecell}
\usepackage{subcaption}
\usepackage{makecell}

\begin{document}
\title{Retrieval-Augmented Approach for Unsupervised Anomalous Sound Detection and Captioning without Model Training}

%


\author{\IEEEauthorblockN{Ryoya Ogura, Tomoya Nishida, Yohei Kawaguchi}
\IEEEauthorblockA{\textit{Research and Development Group, Hitachi, Ltd.}}
}

\maketitle

\begin{sloppy}

\begin{abstract}
This paper proposes a method for unsupervised anomalous sound detection (UASD) and captioning the reason for detection. While there is a method that captions the difference between given normal and anomalous sound pairs, it is assumed to be trained and used separately from the UASD model. Therefore, the obtained caption can be irrelevant to the differences that the UASD model captured. In addition, it requires many caption labels representing differences between anomalous and normal sounds for model training. The proposed method employs a retrieval-augmented approach for captioning of anomalous sounds. Difference captioning in the embedding space output by the pre-trained CLAP (contrastive language-audio pre-training) model makes the anomalous sound detection results consistent with the captions and does not require training. 
Experiments based on subjective evaluation and a sample-wise analysis of the output captions demonstrate the effectiveness of the proposed method.
\end{abstract}
\vspace{-4pt}

\section{Introduction}
\vspace{-5pt}
Unsupervised anomalous sound detection (UASD) is the task of identifying whether sounds emitted by the target machine are normal or anomalous by only using normal sounds as training data \cite{Dohi_arXiv2023_01}. Automatic detection of mechanical failure is essential for artificial intelligence (AI)–based factory automation, as timely detection of machine anomalies via sounds is an effective method for machine condition monitoring.

Various methods for UASD have been explored \cite{8501554,9054344}. Almudevar et al. \cite{almudevar23_interspeech} proposed a variational classifier that outputs an embedding that follows a distribution dependent on the class of the input. Shimonishi et al. \cite{shimonishi23_interspeech} proposed using sound separation for preprocessing to eliminate background noise and unintended sounds from machine sounds. 
However, the proposed systems only distinguish between anomalous and normal sounds without identifying how anomalous sounds deviate from normal ones. Therefore, even when a sound is classified as anomalous, further investigation is required to determine the cause of the machine's anomaly. 

To express the differences between normal and anomalous sounds, Tsubaki et al.\cite{Tsubaki2023} proposed a method that inputs both normal and anomalous sounds and outputs captions describing their differences, which we will refer to as ``difference captions." 
In this method, embeddings of normal and anomalous sounds obtained by an audio encoder, as well as the subtraction of those embeddings, are concatenated to form a single embedding.
This concatenated embedding is then inputted into a decoder that employs a transformer architecture. 
The decoder yields difference captions that articulate the distinctions between normal and anomalous sounds. 
Furthermore, utilizing the MIMII-DG \cite{Dohi2022} dataset, an anomaly was assigned for each normal sound, and the changes were annotated with onomatopoeia, creating the MIMII-Change dataset. This dataset was then utilized to train the proposed model. 
However, this method assumes that difference-caption generation is conducted independently of UASD, which could result in difference captions that do not directly align with the reasons for anomaly detection.
Also, this approach necessitates model training with labeled pairs of normal and anomalous sounds, where preparing the captions for training data and training the model can both be time-consuming.

In this paper, we propose a method that can jointly conduct UASD and caption the differences between normal and anomalous sounds without training on difference-annotated pairs of normal and anomalous sounds. We employ CLAP \cite{CLAP2023} as the backbone: the embedding obtained from the audio encoder of CLAP facilitates UASD, and by inputting this embedding into the text decoder of CLAP, we can generate captions for both normal and anomalous sounds. Then, by comparing the captions of the normal sound and the anomalous sound using GPT-4~\cite{openai2023gpt4}, we can obtain a difference caption. Since CLAP is pre-trained on a large dataset, it does not require additional training for either discrimination or captioning tasks. In our experiments, we first found that CLAP shows sufficient performance on anomalous sound detection. In addition, by observing the difference captions for the sounds identified as anomalous, we were able to obtain reasonably valid captions that matched the cause of the anomalies. 

\begin{figure*}[t]
  \centering
  \includegraphics[keepaspectratio,width=15cm]{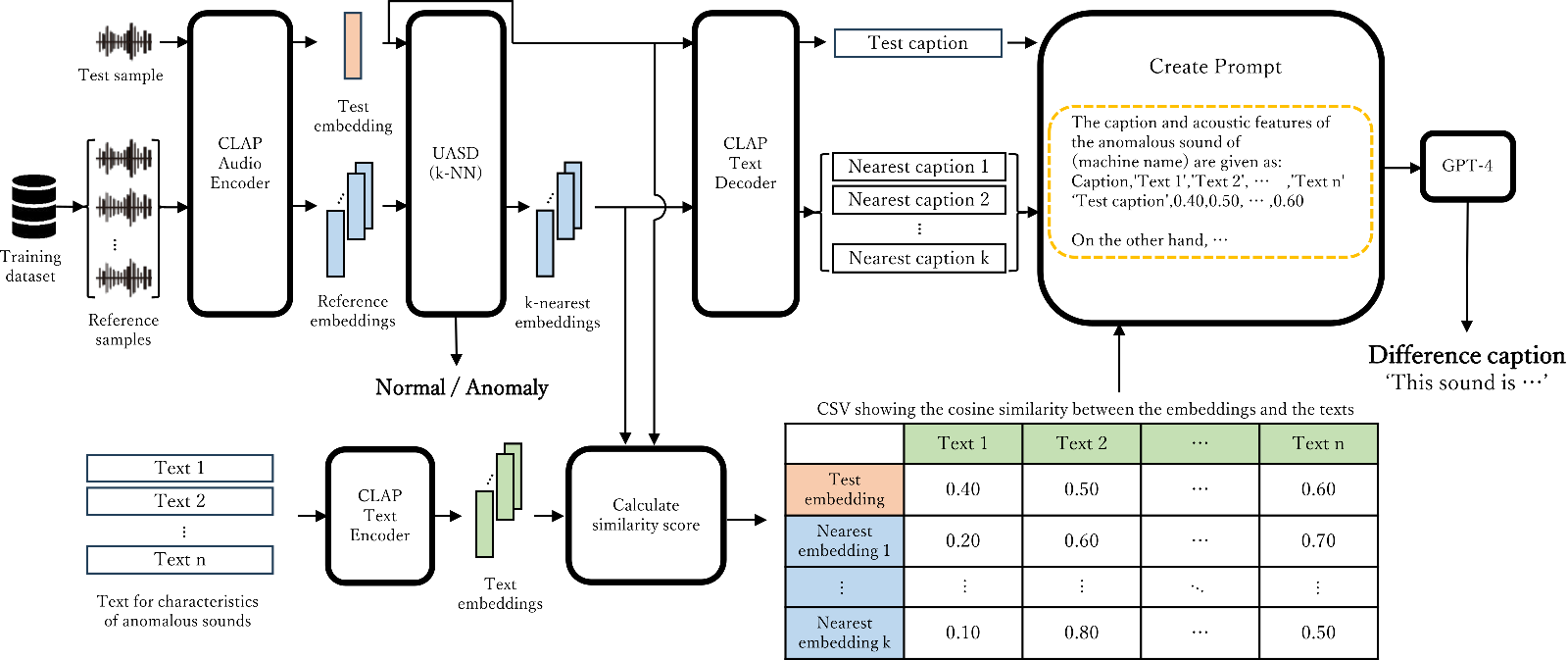}
  \caption{UASD and captioning flow in our proposed method.}
  \label{propsed method}
  \vspace{-10pt}
\end{figure*}

\section{Related Work}
\vspace{-5pt}
\subsection{CLAP}
CLAP is a methodology that links audio and text within a multimodal space. 
Pairs of audio and text are fed into an audio encoder and a text encoder.
The learning process is conducted to ensure that the resulting audio embedding and text embedding match. Utilizing CLAP facilitates zero-shot classification. For example, by calculating the similarity between the embedding of an input audio and the text embeddings of various labels, the most appropriate label for the input exhibits higher similarity. Consequently, this allows for sound event classification without requiring pre-labeled audio for training. 

The experiments in this paper utilize CLAP, which was released by Microsoft in 2023 \cite{CLAP2023}. This CLAP has been pre-trained on a large scale using datasets such as WavCaps \cite{mei2023wavcaps} and AudioSet \cite{7952261}. Audio transformers (HTSAT \cite{9746312}) are used for the audio encoder, and RoBERTa \cite{zhuang-etal-2021-robustly} is used for the text encoder within CLAP. Additionally, the text decoder, consisting of a mapper network and GPT-2 \cite{Radford2019LanguageMA}, can generate sentences from an embedding, which can be combined with the audio encoder to perform audio captioning.

\subsection{Retrieval-Augmented Audio Caption (RECAP)}
RECAP is a method proposed by Ghosh et al. \cite{Ghosh2023RECAPRA} for audio captioning. Audio captioning is the task of describing input environmental sounds in text, and most previous models consist of a pre-trained audio encoder and a text decoder. 
However, these encoder-decoder architectures do not work well when the input data is from a domain different from the trained data.
To address the problem caused by this domain shift, Ghosh uses CLAP to measure the similarity between the input sound and each caption in the data store, and the top four similarity captions are input to GPT-2 to estimate the caption of the input sound. 

RECAP is based on the concept of retrieval augmented generation (RAG) \cite{NEURIPS2020_6b493230}, where information from an external data store is added to the training data for generation. RAG is considered effective in captioning anomalous sounds. 
For example, as pre-trained models such as CLAP are not necessarily trained by the sounds of industrial equipment, generating captions of anomalous machine sounds may not result in captions explaining how the sound is anomalous. 
However, adding additional information, such as captions for normal sounds, allows comparisons between normal and anomalous sounds and can create detailed captions for anomalous sounds.

\section{Proposed Method}
\vspace{-5pt}
\subsection{UASD with CLAP}
\vspace{-5pt}
We use CLAP for UASD. Figure~\ref{propsed method} shows the overview of the proposed method. The training dataset, comprising solely normal sounds, and the test dataset, encompassing both normal and anomalous sounds, are fed into CLAP's audio encoder to derive corresponding embeddings for each dataset. For anomaly detection, the k-Nearest Neighbors (k-NN) is applied. 
The anomaly score $\mathcal{A}(x)$ is defined as
\begin{equation}
 \label{a_score}
\mathcal{A}(x) = \frac{1}{k} \sum_{i=1}^{k} \lVert x - X_{i} \rVert_2,
\end{equation}
where $x$ is the embedding of a test sample. 
$X_i~(i=1,\dots,k)$ are the embeddings of the reference samples from the training database, e.g., $X_i$ is the $i$-th closest embedding in the training database to $x$. 
Anomalous sounds tend to diverge from the distribution of normal sounds, resulting in higher anomaly scores compared to normal sounds in the test dataset. 
Each test data will be classified as normal or anomalous based on a threshold.


\subsection{Difference caption generation for anomalous sounds}
\vspace{-3pt}
The basic idea to generate difference captions without any additional training is to first independently create captions for each anomalous and reference normal sounds using an existing text decoder, and then let large language models, GPT-4 in our experiments, compare those output captions.
To create captions of each sounds, we utilize the CLAP embeddings used in UASD.
The embedding of the sound identified as anomalous in UASD, $X^{(a)}$, is input to the CLAP's text decoder provided for the model in \cite{CLAP2023} to generate a caption that explains this sound.
Embeddings of the k reference samples $X_1,\dots X_k$ are also fed into the same text decoder to generate captions for k instances of normal sounds.
All these $k+1$ captions are then used to form a prompt that asks GPT-4 how the caption of the anomaly-identified sound is different from the other captions.
Specifically, the prompt is formed as {\it ``The caption of the anomalous sound of [machine name] is given as: [caption of the anomaly-identified sound]. On the other hand, the captions of the normal sounds of [machine name] are given as: [k captions of the normal sounds, separated with commas]. Please describe in broad strokes how this anomalous sound differs compared to the normal sounds."}
The output of GPT-4 is then directly used as the difference caption which explains how the anomaly-identified sound is different from sounds in the normal training data.
In this paper, we will call this method the ``text decoder-based method".

Since this method uses the same embeddings for UASD and for generating difference captions, it is expected that the difference captions will not contradict the UASD results.
In addition, since we only use pre-trained models as it is, additional learning for UASD or difference caption generation is unnecessary.
Note that this also means that the output difference caption depends on CLAP's captioning performance.
Appropriate difference captions are expected to be obtained for cases where the anomalous and normal sounds has differences that CLAP's captioning network can capture.
We assume that such a case holds for some typical anomalous conditions, since anomalous conditions can often cause additional sounds that have different characteristics than the original normal sounds, e.g., rattling sounds added to smooth movement sounds, or cause certain sounds disappear, which is likely to appear in the captions.


\subsection{Difference caption generation with predefined texts}
\vspace{-3pt}
While certain differences between the anomalous and normal sounds can be captured by the first method, in some cases, CLAP's text decoder can provide very similar descriptions for the anomalous sound and the k reference normal sounds to be compared.
This can happen even when the distance between the embeddings are large and the anomaly score is high.
In such cases, comparing those descriptions in GPT-4 cannot acquire appropriate captions that explain the differences between anomalous and normal sounds.

To solve this problem, we additionally propose a method based on the zero-shot classification framework using CLAP. 
First, a set of $L$ short reference texts $text_1,\dots, text_L$, describing common characteristics of malfunctioning machine sounds, such as ``Vibration" and ``Popping and Knocking sounds" are prepared. 
These texts can for example be prepared by asking large language models.
For each of these texts, the text embeddings $T_l~(l=1,\dots, L)$ are extracted through CLAP's text encoder.
Then, the cosine similarities $S_C(X, T_l)$ between each text embedding $T_l$ and audio embedding $X \in \{X^{(a)}, X_1,\dots,X_L\}$ are computed to infer how much that audio input contains the feature described in the $l$th text.
Finally, the obtained similarity scores are included into a prompt that asks to compare those values for the anomaly-identified sound and the normal reference sounds.
Specifically, the prompt is formed as {\it ``The acoustic features of the anomalous sound of [machine name] are given as [similarity scores for the anomaly-identified sound]. On the other hand, the acoustic features of the k normal sounds of [machine name] are given as [similarity scores for the $k$ reference normal sounds]".}
Here, the similarity scores are given as a CSV format where the rows represent scores for each reference text and the columns represent scores for each sound, e.g., for the reference normal sounds, the scores are given as 
{\footnotesize
\begin{align}
    &text_1,~ text_2,~ \dots,~ text_L,~ \backslash\text{n}~ \notag\\
    &S_C (X_1, T_1),~ S_C (X_1, T_2),~ \dots,~ S_C (X_1, T_L),~ \backslash\text{n} \notag\\
    &\dots,~ \notag\\
    &S_C (X_k, T_1),~ S_C (X_k, T_2),~ \dots,~ S_C (X_k, T_L),~\notag
\end{align}
}
If the similarity between some reference text and the sound is different for the normal sounds and the anomaly-identified sound, that difference should be extracted as the final output caption.
We will call this method the ``Zero-shot classification-base method".

This strategy can also be combined with the method proposed in the previous section.
This is realized by including both the caption information and the similarity scores is one CSV format, such as 
{\footnotesize
\begin{align}
    &\text{caption},~ text_1,~ text_2,~ \dots,~ text_L,~ \backslash\text{n} \notag\\
    &\text{[output caption]},~ S_C (X_1, T_1),~ S_C (X_1, T_2),~ \dots,~ S_C (X_1, T_L),~ \backslash\text{n} \notag\\
    &\dots,~ \notag
\end{align}
}%
In this way, the output difference caption would include both caption based and specified characteristics-based explanations.

\section{Experiment}
\vspace{-5pt}
\setlength{\tabcolsep}{1mm}
\begin{table}[t]
\begin{center}
\vspace{-10pt}
\caption{AUC(\%) for baseline and pre-trained models.}
\label{tab:auc_results}
\vspace{-5pt}
\scriptsize
\scalebox{0.9}{
\begin{tabular}{@{}c | c | c c c c@{}}
\hline
\ \\[-6.5pt]
Machine & ID & Baseline & PANNs & LAION-CLAP & MS-CLAP \\ \hline\hline
     & 01 & {\bf 81.36} & 80.04 & 76.94 & 75.68  \\
     & 02 & {\bf85.97} & 85.92 & 84.11 & 80.99  \\
ToyCar & 03 & 63.30 & {\bf69.98} & 64.76 & 65.42  \\
     & 04 & 84.45 & {\bf89.42} & 85.53 & 83.57  \\ \Xcline{2-6}{0.1pt}
     & Average & 78.77 & {\bf81.34} & 77.84 & 76.42  \\ \hline
     & 01 & {\bf78.07} & 61.65 & 63.14 & 63.42  \\
     & 02 & {\bf64.16} & 56.45 & 55.78 & 59.23  \\ 
ToyConveyor & 03 & {\bf75.35} & 60.24 & 58.28 & 59.65  \\ \Xcline{2-6}{0.1pt}
     & Average & {\bf72.53} & 59.45 & 59.07 & 60.77   \\ \hline
     & 00 & {\bf54.41} & 50.63 & 53.87 & 51.91  \\
     & 02 & {\bf73.40} & 60.25 & 65.21 & 61.89  \\
Fan  & 04 & {\bf61.61} & 45.77 & 47.78 & 46.88  \\
     & 06 & 73.92 & {\bf75.21} & 74.40 & 67.96  \\ \Xcline{2-6}{0.1pt}
     &Average & {\bf65.83} & 57.97 & 60.32 & 57.16  \\ \hline
     & 00 & 67.15 & 81.92 & {\bf87.40} & 84.22  \\
     & 02 & 61.53 & 68.23 & 89.13 & {\bf92.49}  \\
Pump & 04 & {\bf88.33} & 77.71 & 83.27 & 80.39  \\
     & 06 & {\bf74.55} & 56.83 & 62.78 & 58.83  \\ \Xcline{2-6}{0.1pt}
     & Average & 72.89 & 71.17 & {\bf80.64} & 78.98  \\ \hline
     & 00 & 96.19 & 99.89 & {\bf99.98} & 99.25  \\
     & 02 & 78.97 & 88.58 & 91.25 & {\bf93.87}  \\
Slider & 04 & {\bf94.30} & 86.51 & 79.21 & 80.46  \\
     & 06 & 69.59 & 57.31 & 61.99 & {\bf71.28}  \\ \Xcline{2-6}{0.1pt}
     & Average & 84.76 & 83.07 & 83.11 & {\bf86.22}  \\ \hline
     & 00 & 68.76 & 84.38 & {\bf91.07} & 80.53  \\
     & 02 & 68.18 & {\bf79.89} & 79.39 & 76.61  \\
Valve & 04 & 74.30 & 79.26 & 80.80 & {\bf81.45}  \\
     & 06 & 53.90 & 69.53 & 64.22 & {\bf70.23}  \\ \Xcline{2-6}{0.1pt}
     &Average & 66.28 & 78.26 & {\bf78.87} & 77.21  \\ \hline
\end{tabular}
}
\vspace{-7pt}
\end{center}
\end{table}

\begin{figure}[t]
    \center
    \includegraphics[width=0.55\linewidth,clip]{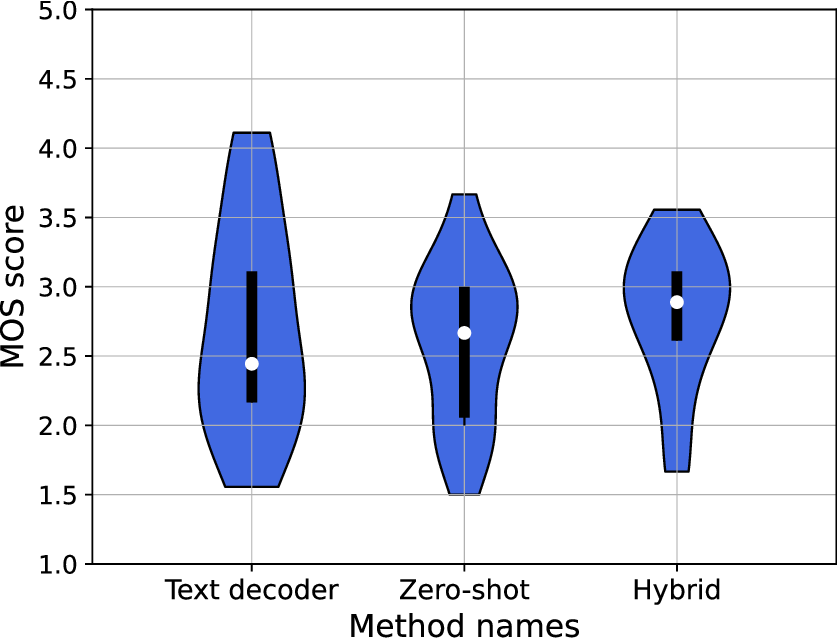}
    \vspace{-5pt}
    \caption{Violin plot for MOS values of each ID in each machine. "Text decoder``, "Zero-shot``, and "Hybrid`` denote Text decoder-based method, Zero-shot classification-based method, and combination of these methods, respectively. White dot denotes the median and black bar denotes range of quartiles.}
    \vspace{-12pt}
    \label{fig:mos}
\end{figure}

\begin{figure*}[t]
    \center
    \includegraphics[width=0.85\linewidth,clip]{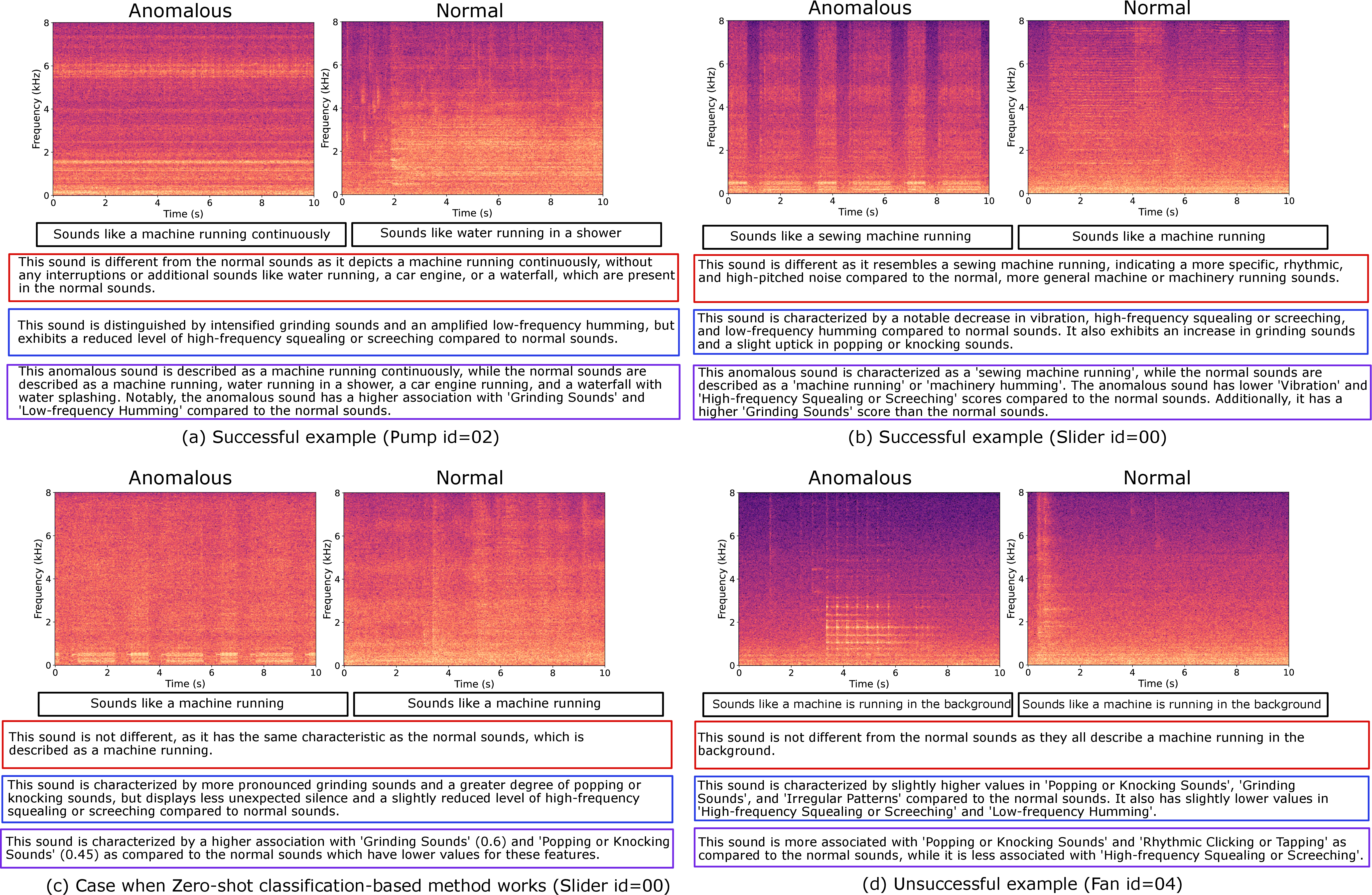}
    \vspace{-5pt}
    \caption{Spectrograms of anomalous and reference normal sounds and corresponding captions. Black box (first row) denotes individual captions for each sample; Red box (second row) denotes captions for CLAP's text decoder-based method; Blue box (third row) denotes captions for zero-shot classification-based method; Purple box (forth row) denotes captions for combination of both methods.}
    \vspace{-6pt}
    \label{fig:specs}
\end{figure*}



\subsection{Experimental conditions}
\vspace{-3pt}
We evaluated the proposed method using an ASD dataset. 
To verify the effectiveness of the proposed method under simple conditions where no domain shifts occur, we used the DCASE 2020 Challenge Task 2 Development Dataset \cite{Koizumi_DCASE2020_01} for the experiments. 
It contains four types of machines (Fan, Pump, Slide rail, Valve) from MIMII Dataset \cite{Purohit_DCASE2019_01} and two types of toys (Toy-car, Toy-conveyor) from ToyADMOS \cite{Koizumi_WASPAA2019_01} with normal and anomalous sounds.

In this experiment, we set $k=4$. The performance of UASD per Machine ID is calculated using the area under the receiver operating characteristic curve (AUC). The CLAP audio encoder by Microsoft used in our proposed method was compared in performance with pre-trained models such as PANNs \cite{9229505} and the CLAP audio encoder by LAION \cite{laionclap2023}, in addition to the autoencoder used as a baseline for the DCASE 2020 Challenge Task 2\cite{Koizumi_DCASE2020_01}.

Next, captions were generated for the test data identified as anomalous sounds in the UASD. Three methods of generation were used: the text decoder-based method, the zero-shot classification-based method, and the combination of both methods. 
For all methods, a common prompt instructing the GPT-4 to begin their output sentence with ``This sound is" and to finish it within 40 words was added before each prompt described in the previous section.
For the text decoder-base method and the combined method, we constrained the text output from CLAP's text decoder to be in the form of ``Sounds like \dots".
We used the following 8 texts as descriptions of anomalous sounds for the zero-shot classification-based method: ``Vibration", ``High-frequency Squealing or Screeching", ``Popping or Knocking Sounds", ``Rhythmic Clicking or Tapping", ``Grinding Sounds", ``Irregular Patterns", ``Low-frequency Humming" and ``Unexpected Silence".
These texts were created by asking GPT-4 to provide some common characteristics of the sounds of malfunctioning machines.

We subjectively evaluated the output captions by the Mean Opinion Score (MOS).
For evaluation, we selected three samples for each ID of each machine type that showed the highest anomaly scores, since captions for high anomaly score samples matters the most.
In total, $69$ sets of data-caption pairs were evaluated for each method.
Three non-expert subjects were asked to listen to the test and reference normal samples, and rate how well each caption explains the differences between them.
The ratings ranged from '1' to '5', where '1' represents the worst and '5' represents the best.
In addition, we reviewed each created caption to verify if they align with the cause of the anomaly and sound changes that should occur from that anomaly.

\subsection{UASD experimental results}
Table~\ref{tab:auc_results} shows the experimental results of UASD using CLAP embeddings along with the baseline and other pre-trained models. 
``MS-CLAP" refers to the CLAP by Microsoft.
The results indicate that the MS-CLAP's embedding achieves a performance level almost parallel to that of other models, with the performance variance between CLAP and the other models dependent on the specific model and Machine ID.
This discrepancy in performance may be attributed to the slight differences in the dataset used for pre-training the models.
The findings also suggest that by applying transfer learning and fine-tuning to CLAP's audio encoder, UASD performance could potentially be enhanced further, leveraging the pre-trained model's capabilities more effectively.

\vspace{-4pt}
\subsection{Captioning experimental results}
\vspace{-4pt}
Fig.~\ref{fig:mos} shows the distribution of the MOS values computed for each ID in each machine.
Overall, the MOS values for a considerable portion of IDs were over '3', indicating that the output captions can be informative to some extent.
The number of IDs that received MOS values over '3' was $10$, $8$, and $11$ out of $23$, for the text-decoder-based method, the zero-shot classification-based method, and the combination method, respectively.
The Text-decoder-based method showed significantly higher MOS values for several IDs than other methods, such as pump id00 (4.1), slider id00 (4.0), and ToyCar id01 (3.8), indicating its usefulness for certain kinds of data.
At the same time, a large portion of IDs had relatively low MOS values such as between $1.5$ to $2.5$ for this method.
This was mainly because the text decoder sometimes provided very similar captions for both anomalous and normal sounds, resulting in difference captions that are not very expressive.
On average, the combination of the two proposed methods showed the highest MOS value, which indicates this method was able to take the good parts of both methods to some extent.

Fig.~\ref{fig:specs} shows examples of the difference captions generated by the three methods against anomalous test samples. 
We picked typical examples to show the general trends of the proposed methods.
(a) shows an example for pumps, where the text decoder-based method describes the absence of water sounds in the anomalous sound.
This coincides with the cause of malfunction in pumps, where pumps in normal conditions have water flows but anomalous condition does not, due to leakage or clogging~\cite{Purohit_DCASE2019_01}.
The zero-shot classification-based method generated the difference caption focusing on the low frequency of the anomalous sound, which can also be seen in the spectrograms.
(b) shows an example for sliders, which is also a successful example. 
In this case, an anomalous slider makes sounds caused by scrapings or rattlings of rails due to rail damage or lack of grease~\cite{Purohit_DCASE2019_01}. 
Expressions such as ``sewing machine" and ``grinding" match such characteristics, explaining strong periodic sounds or sounds related to hard substances scraping each other.
(c) is another slide rail example, where the text decoder-based method did not capture any differences. 
However, the zero-shot classification-based method successfully described the difference, explaining the difference in ``grinding" sounds.
This shows the effectiveness of the zero-shot classification-based method, even when the text decoder-based method does not work ideally.
(d) is an example for fans, where none of the methods successfully described the difference. 
While the zero-shot classification-based method described differences such as higher ``Popping or Knocking sounds" or ``Irregular Patterns", they seem to just coincide with the background noise that appear in 3.5(s)--6 (s) in the anomalous sound.
As shown in this example, obtaining appropriate difference captions for fans seemed difficult, which is reasonable considering that the AUC values for fans were low.

\vspace{-2pt}
\section{Conclusion}
\vspace{-5pt}
In this paper, we proposed a method for jointly conducting UASD and difference captioning based on CLAP embedding. 
The use of CLAP embedding confirmed that UASD is possible with high performance. 
In addition, results indicated that the RAG-based method for difference-caption generation can generate captions that match why the samples are detected as anomalous to some extent.
Furthermore, the proposed method does not require the time for training in conventional methods because the proposed method only uses a pre-training model.

\bibliographystyle{IEEEtran}
\bibliography{mybib}

\end{sloppy}
\end{document}